# Temporal metastates are associated with differential patterns of time-resolved connectivity, network topology, and attention

James M. Shine[a,b,1], Oluwasanmi Koyejo[a], and Russell A. Poldrack[a]

[a]Psychology Department, Stanford University, Stanford, CA 94301; and [b]Neuroscience Research Australia, The University of New South Wales, Sydney, NSW 2031, Australia

**Little is currently known about the coordination of neural activity over longitudinal timescales and how these changes relate to behavior. To investigate this issue, we used resting-state fMRI data from a single individual to identify the presence of two distinct temporal states that fluctuated over the course of 18 mo. These temporal states were associated with distinct patterns of time-resolved blood oxygen level dependent (BOLD) connectivity within individual scanning sessions and also related to significant alterations in global efficiency of brain connectivity as well as differences in self-reported attention. These patterns were replicated in a separate longitudinal dataset, providing additional supportive evidence for the presence of fluctuations in functional network topology over time. Together, our results underscore the importance of longitudinal phenotyping in cognitive neuroscience.**

Methodological advances in cognitive neuroscience have enabled increasingly intricate descriptions of neural dynamics using fMRI (1). Studies leveraging these techniques have highlighted a set of large-scale cortical networks (2) that are among the most flexible (3) and dynamic (4) regions in the brain (5). Additional work in the field has shown that coordinated and adaptable patterns of functional connectivity between these regions underpin a number of higher brain functions, such as cognition (3), learning (6), and consciousness (7). This work has largely focused on the study of individuals at single time points, but it is clear that there are also changes in brain connectivity over much longer timescales of weeks to months (8, 9). Importantly, it is not currently known how these long-term changes in connectivity are related to momentary dynamic changes and how these time-resolved patterns are related to psychological function.

Here, we leveraged a unique longitudinal resting-state fMRI (rfMRI) dataset (8) to determine whether fluctuations in whole-brain connectivity were associated with alterations in the dynamic organization of the resting brain over the course of weeks to months. First, we used affinity propagation to cluster the time-averaged connectivity patterns from 84 separate rfMRI scanning sessions, which revealed the presence of two intermittently present "metastates" (Fig. 1). Second, we then used the multiplication of temporal derivatives (MTD) (10) technique to estimate patterns of time-resolved functional connectivity within each session. By tracking the community structure of the brain within 10-s windows over the course of each scanning session, we were able to estimate both global and local patterns of time-resolved connectivity (11). To estimate time-resolved connectivity at the areal level, we used a previously described measure of network-level interareal dynamic connectivity—"flexibility"—which describes the extent to which a given brain region switches frequently between distinct communities over time (12).

## Results

Over the course of 18 mo, a single individual (R.A.P.; male; age 45 y old at the onset of the study) underwent 104 scanning sessions, of which 84 had rfMRI data suitable for subsequent analysis (8). Time series were extracted from a series of 630 cortical and subcortical parcels (13), which were then used to create a 630 × 630 parcelwise time-averaged correlation matrix for each individual scanning session (Fig. 1*A*). Affinity propagation clustering (14) identified two major clusters (Fig. 1*B*), confirming the existence of two large metastates that intermittently fluctuate over longitudinal time (Fig. 1*C*), significantly more frequently than would be predicted by a stationary null model ($P < 0.001$). Importantly, there were no differences in head motion [as measured by mean framewise displacements (MFDs)] between the two states ($MFD_1 = 0.106 \pm 0.01$; $MFD_2 = 0.109 \pm 0.12$; $P > 0.200$).

Next, we investigated the time-resolved connectivity between parcels within each session by calculating the pointwise product of the temporal derivative of each time series (MTD) (10) within a sliding window of 10 s. The MTD, which is conceptually similar to a sliding window correlation of temporally differentiated time series, has previously been shown to show improvements in sensitivity to shifts in connectivity structure compared with sliding window Pearson's correlations of undifferentiated time series. The MTD is also less susceptible to known sources of spurious connectivity, such as head motion and global mean signal fluctuations (10). The calculation of the MTD enabled the

### Significance

The brain is an inherently dynamic organ; however, the manner in which the brain changes over longitudinal time remains poorly understood. An understanding of these dynamic mechanisms is critical for understanding normal childhood development and aging as well as neurological and psychiatric disease states. Here, we leverage data collected in a single individual over the course of 2 y to investigate changes in brain organization over time. In doing so, we show intermittent fluctuations in brain network configuration that were associated with separable patterns of time-resolved interactions between neural regions. The states also directly relate to alterations in whole-brain information processing and self-reported attention. In addition, the patterns that we observed were replicated in a separate longitudinal dataset.

Author contributions: J.M.S. and R.A.P. designed research; J.M.S. and O.K. analyzed data; and J.M.S., O.K., and R.A.P. wrote the paper.

The authors declare no conflict of interest.

This article is a PNAS Direct Submission.

Freely available online through the PNAS open access option.

[1]To whom correspondence should be addressed. Email: macshine@stanford.edu.

This article contains supporting information online at www.pnas.org/lookup/suppl/doi:10.1073/pnas.1604898113/-/DCSupplemental.

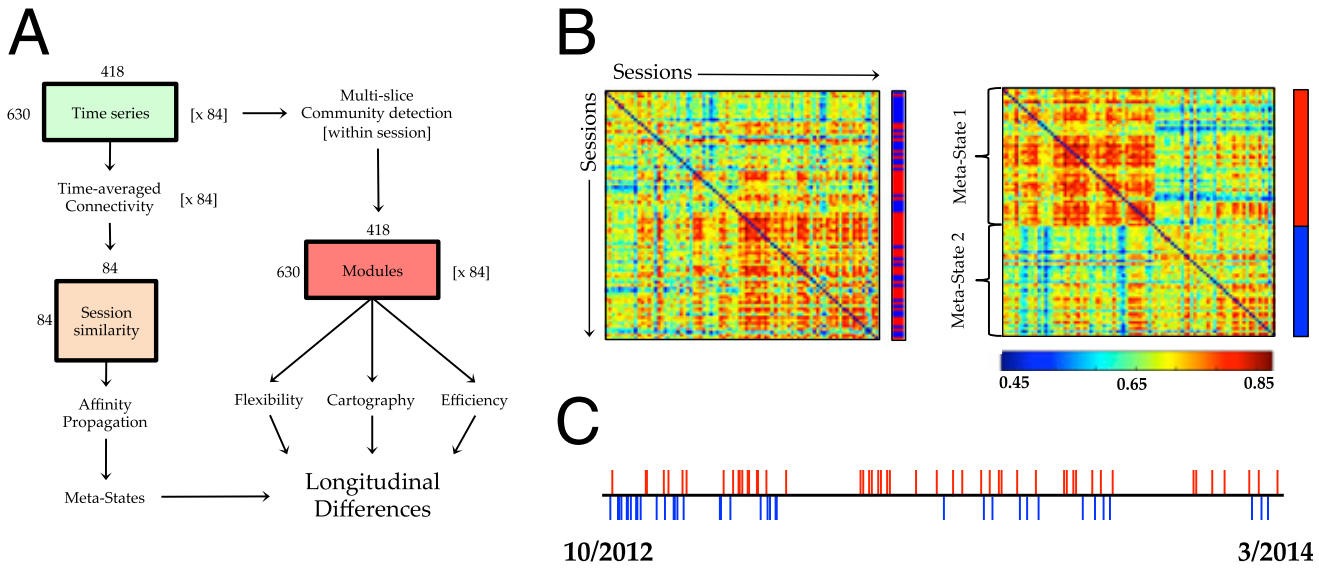

**Fig. 1.** (*A*) Graphical depiction of the experiment—time series from 630 cortical and subcortical parcels in 84 separate sessions were submitted to time-averaged connectivity analysis. Affinity propagation was then used to cluster the similarity of each session's time-averaged connectivity. Separately, the time series from each session were subjected to a time-resolved functional connectivity analysis, and then, a multislice community detection algorithm was used to track the modular structure of the brain over time. (*B*) Spatial similarity of parcelwise resting-state functional connectivity matrices for each session over time [the cluster identity of each session is represented as either red (metastate 1) or blue (metastate 2) in the vector alongside the adjacency matrix]—there were two temporal metastates identified at the group level using affinity propagation clustering of the similarity between time-averaged connectivity matrices (metastate identity shown alongside adjacency matrix). (*C*) A timeline showing the relative occurrence of each session colored according to its metastate. A similar pattern was observed in the replication dataset (Fig. 4*E*).

estimation of a parcel × parcel × time adjacency matrix for each of the 84 individual scanning sessions. The 3D matrix for each session was then subjected to a multislice modularity analysis (15), which estimates the presence of communities of brain regions that extend over time. The community assignment within each temporal window was then used to estimate the within- ($W_T$) and between-module ($B_T$) connectivity for each region (*SI Materials and Methods*). At the whole-brain level, the tradeoff between $W_T$ and $B_T$ (the cartographic profile) can be tracked over time, thus providing an estimate of temporal fluctuations in system-wide integration and segregation (11).

At the areal level, we reasoned that regions important for dynamic communication should show "flexible" behavior (6)—that is, a dynamic region should communicate with a wide variety of regions over time and hence, switch between modules frequently over the course of a resting session. As an initial step, we estimated the flexibility of each of 347 regions (333 cortical and 14 subcortical) within a single rfMRI session [repetition time (TR) = 0.72 s; spatial resolution = 2 mm$^3$] from 100 unrelated individuals from the Human Connectome Project (HCP) (16). We observed marked heterogeneity in the flexibility of neural regions in the resting state, with regions within default and salience network, along with a number of subcortical regions, showing the most flexible behavior during rest. In contrast, regions within the frontoparietal network showed the most stable behavior during rest (Fig. 2*B*). We also observed similar patterns of flexibility across 84 sessions from the individual subject and a similar longitudinal dataset from a single individual ("Kirby") (Fig. 2*C*) (8) as well as in data from 100 unrelated individuals from the HCP Consortium (spatial correlation between mean regional flexibility from 100 subjects in HCP and mean flexibility across 84 sessions in the MyConnectome Project: $r = 0.440$) (Fig. 2*A*), suggesting that the flexibility of brain regions over time is relatively stable across subjects and datasets.

We were next interested in determining whether the two temporal metastates showed unique dynamic signatures within the individual resting-state sessions. Indeed, the two metastates were associated with distinct patterns of time-resolved connectivity, with metastate 2 highlighted by markedly increased flexibility in the visual, somatomotor, frontoparietal, and cingulo-opercular networks (Fig. 3*A*) [false discovery rate (FDR) α = 0.05]. These differences (in all but six parcels: left insula, bilateral superior frontal gyrus, and bilateral temporal pole) were significantly greater than the 95th percentile of a null distribution populated by results obtained from a phase randomized dataset (17), which scrambles dynamic interrelationships in the data. In addition, although there were similar patterns of modularity in both states (i.e., the extent to which the network was partitioned into tight knit communities; $Q_1 = 0.609 \pm 0.07$; $Q_2 = 0.600 \pm 0.08$; $P = 0.240$), we observed higher global efficiency (i.e., the ease with which a pair of regions within the largest connected component of the network can communicate; $E_1 = 0.308 \pm 0.02$; $E_2 = 0.319 \pm 0.02$; $P = 0.002$; greater than 95th percentile of phase-randomized null distribution) (Fig. 3*C*) and systems-level integration (i.e., the extent to which the community structure of the brain was dissolved; greater than 95th percentile of phase-randomized null distribution) (Fig. 3*B*) in metastate 2. Together, these results show that the dynamic interplay between frontoparietal and sensorimotor regions is related to differences in the capacity of the whole brain to alter its information processing capacity over longitudinal time.

Given that behavioral capacities, such as attention and alertness, are known to fluctuate over time, we next investigated whether fluctuations in time-resolved connectivity were related to fluctuations in psychological function. To do so, we identified questions from the self-reported Positive and Negative Affect Schedule (18) that were significantly different when collected on days associated with scanning sessions that were later identified as occupying either of the metastates. This analysis revealed a differential relationship (Mann–Whitney $u$ test; FDR $P < 0.05$) between the behavior associated with one of two states, with the less flexible state (metastate 1) corresponding to questions associated with fatigue (drowsy: $Q_{28}$; sleepy: $Q_{57}$; sluggish: $Q_{58}$; and tired: $Q_{62}$) and the state with more flexible interareal dynamics (metastate 2) associated with heightened attention (attentive: $Q_{11}$; concentrating: $Q_{18}$; and lively: $Q_{43}$). Interestingly, the four "fatigue"-related questions were also found to be significantly different when comparing sessions acquired with or without caffeine/food, a factor that was manipulated in the study (8) (all $P < 0.002$). However, caffeine and food were not associated with any of the questions tracking self-reported attention and were similarly not associated with the presence of either metastate (all $P > 0.200$) (8). As such, this finding suggests that the fluctuations in flexibility were not simply related to drowsiness within the scanner; however, definitive resolution of this issue would require simultaneous EEG/fMRI data to track the electrophysiological signatures of sleep architecture (19, 20).

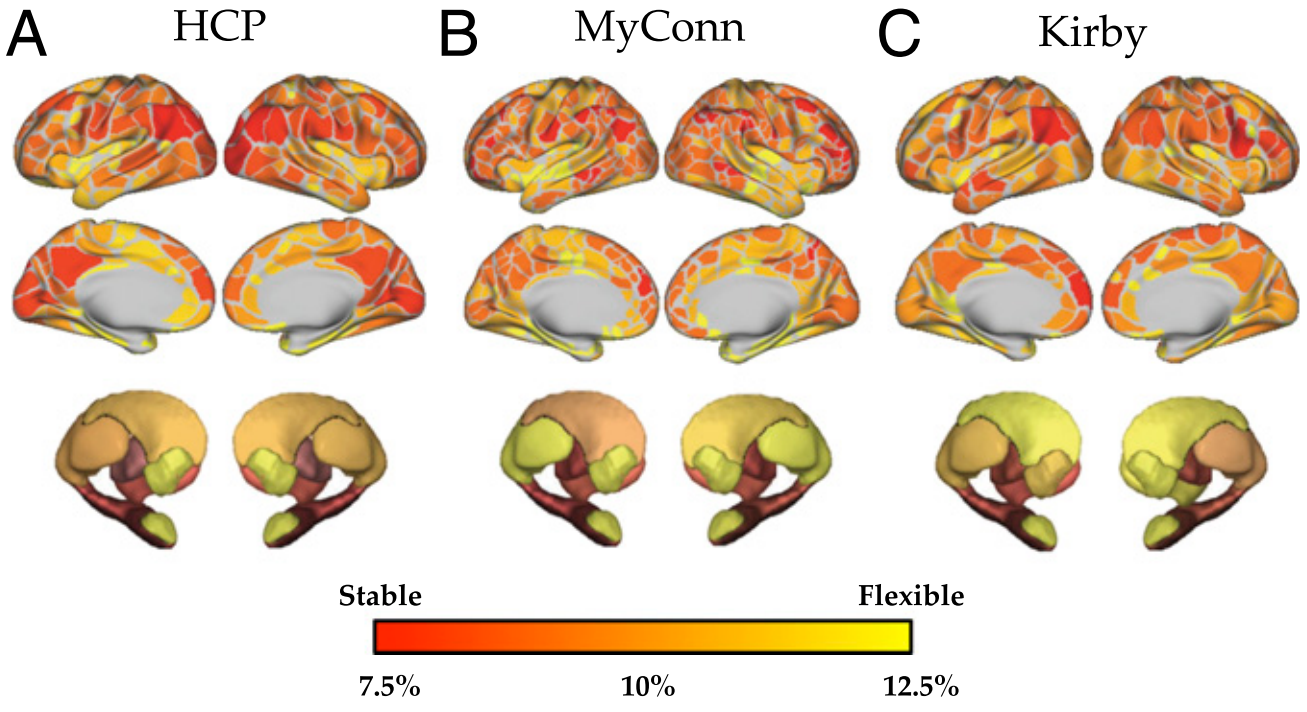

**Fig. 2.** The flexibility (percentage of time "switching" between unique modules) of brain regions across (*A*) 100 unrelated subjects from the HCP, (*B*) 84 sessions from the MyConnectome Project dataset (MyConn), and (*C*) 152 sessions from the Kirby dataset.

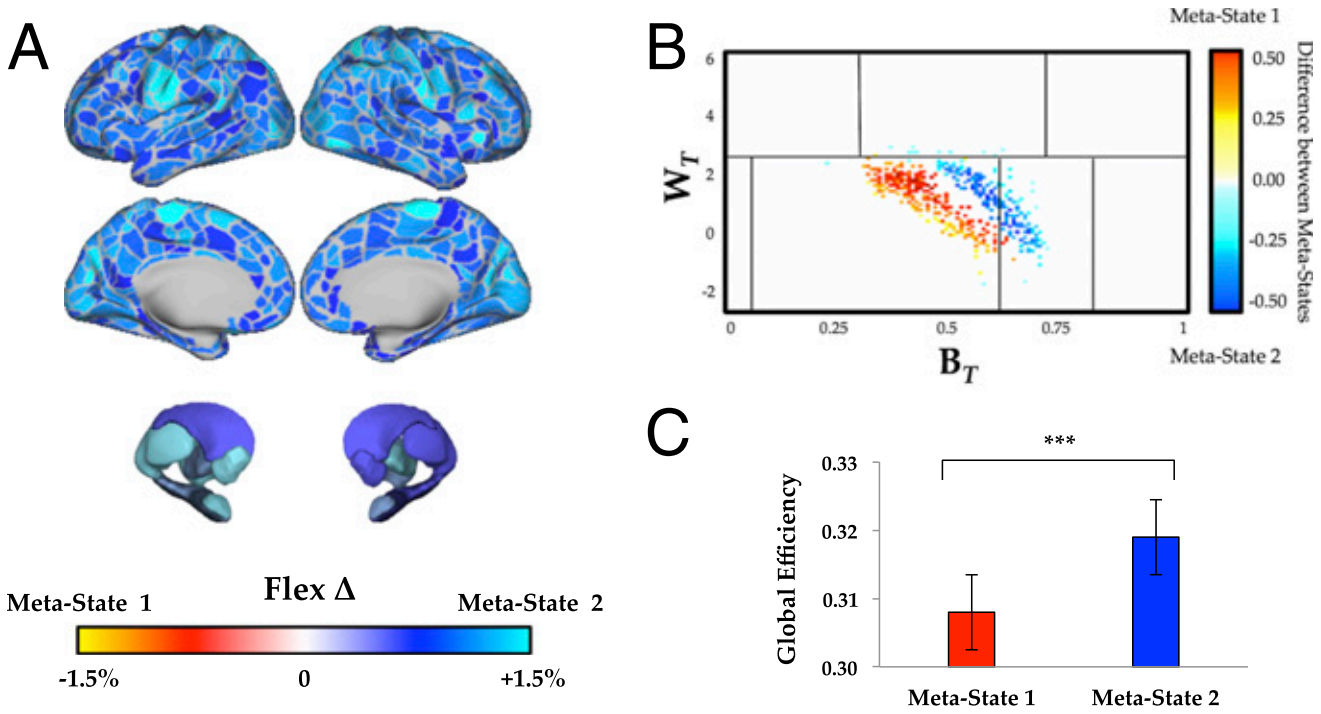

**Fig. 3.** (*A*) Significant differences in patterns of flexibility between the two temporal metastates (FDR $P < 0.05$)—metastate 2 (blue) was associated with higher flexibility than metastate 1 (yellow/red). (*B*) Differences in the cartographic profile between metastate 1 (yellow/red) and metastate 2 (blue)—metastate 2 was associated with a shift toward higher integration (FDR $P < 0.05$). (*C*) Metastate 2 was associated with greater global efficiency than metastate 1 ($P = 0.002$). $^{***}P < 0.01$.

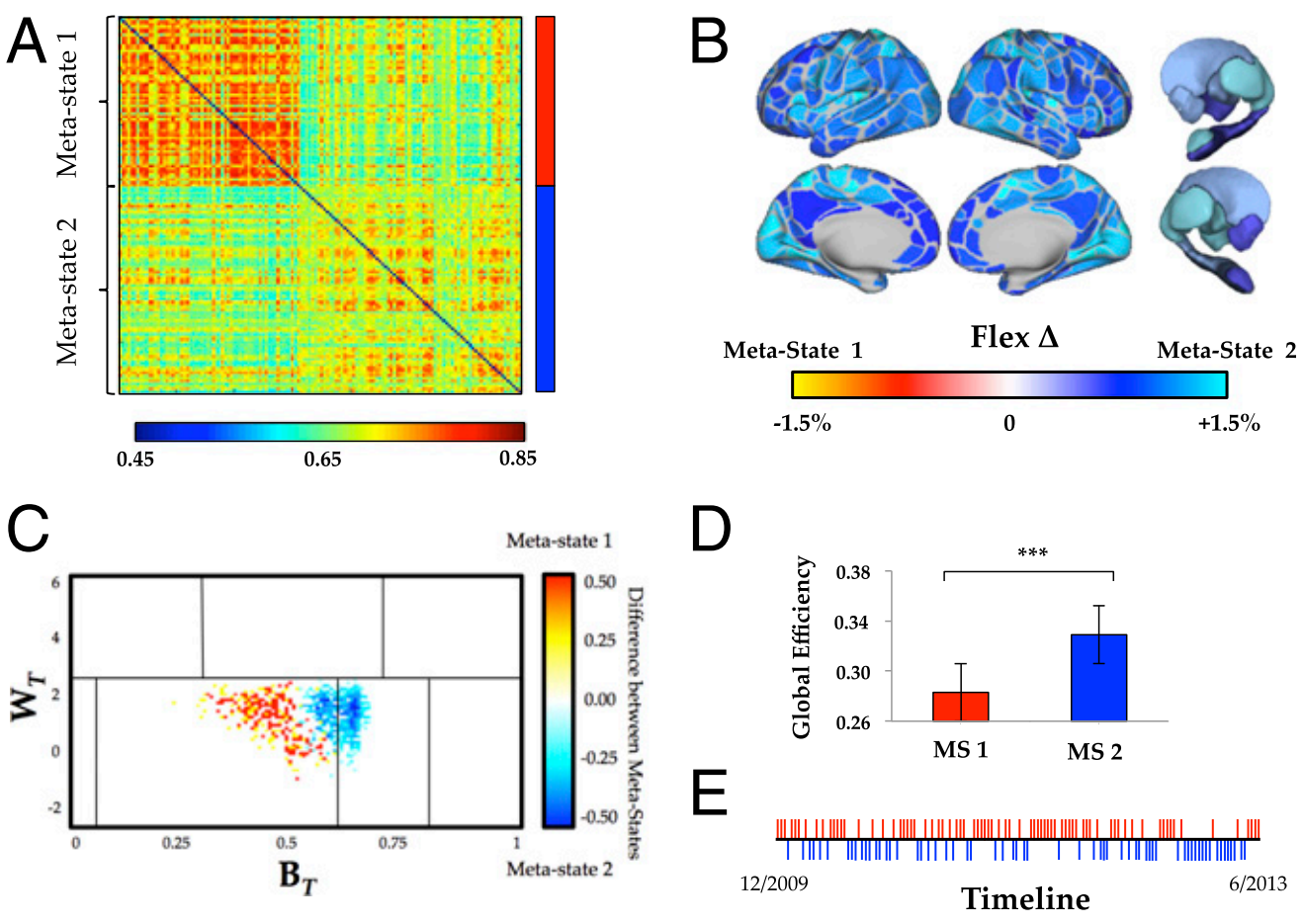

**Fig. 4.** (*A*) We identified two temporal metastates using affinity propagation clustering (replication metastate identity shown alongside adjacency matrix). (*B*) Significant differences in patterns of flexibility between the two temporal metastates (FDR $P < 0.01$): blue, significantly higher time-resolved connectivity in replication metastate 2; red, significantly higher time-resolved connectivity in replication metastate 1. (*C*) Differences in the cartographic profile between replication metastate 1 (yellow/red) and replication metastate 2 (blue)—replication metastate 2 was associated with a shift toward higher integration (FDR $P < 0.05$). (*D*) Replication metastate 2 was associated with greater global efficiency than replication metastate 1 ($P = 0.001$). $^{***}P < 0.01$. (*E*) A timeline showing the relative occurrence of each session colored according to its metastate.

To ensure that the results that we observed were not reflective of idiosyncratic patterns within the MyConnectome Project dataset (e.g., the patterns related to caffeine and food intake) (8), we replicated our analysis in a separate longitudinal dataset of a single individual (www.nitrc.org/projects/kirbyweekly). Briefly, this dataset included 158 sessions collected over 4.5 y in a single individual (male; 40 y old) using a scanning protocol with lower spatial and temporal resolution (TR = 2 s; spatial resolution = 3 $mm^3$). After preprocessing, 138 scanning sessions from this dataset passed quality assessment and were subjected to our analysis. In addition, all data were minimally "scrubbed" to remove the potential effect of head motion on connectivity measures (21, 22), however the results were independent of this preprocessing step. We found similar fluctuations in connectivity (Fig. 4), with two metastates fluctuating intermittently over the period of the study ($Rep_1 = 44.9\%$; $Rep_2 = 55.1\%$) (Fig. 4*E*). Consistent with the findings in the discovery cohort, the two metastates were associated with similar differences in flexibility ($r = 0.360$) (Figs. 4*B* and 5), cartography ($r = 0.536$) (Fig. 4*C*), and network topology [i.e., different global efficiency ($Rep_{E1} = 0.283 \pm 0.05$; $Rep_{E2} = 0.329 \pm 0.04$; $P = 0.002$) but similar modularity ($Rep_{Q1} = 0.545 \pm 0.03$; $Rep_{Q2} = 0.539 \pm 0.02$; $P = 0.312$) (Fig. 4*D*)]. No psychological data were available for this dataset (9), and therefore, the behavioral analyses could not be replicated. Despite this idiosyncracy, the results of the replication analysis provide confirmatory evidence for the presence of metastates over the course of weeks to months.

## Discussion

Here, we showed that fluctuations in rfMRI functional connectivity over the course of weeks to months in a single individual are related to specific patterns of time-resolved connectivity, network topology, and self-reported attention. By tracking large-scale descriptions of functional connectivity over a period greater than 1 y, we identified the presence of two longitudinal metastates that fluctuated over time (Fig. 1 *B* and *C*). These metastates were characterized by separable patterns of time-resolved connectivity at both the global and areal levels. Specifically, we observed quantitative differences in the regions that showed flexible behavior over time, sharing community structure with many regions, while also regularly switching modular assignments (Fig. 3). These findings were replicated across two separate individuals (both with unique scanning protocols), providing some degree of support for their generalizability. However, despite the relatively similar patterns of flexibility associated with each of two metastates (spatial correlation between MyConnectome Project and Kirby datasets: $r = 0.360$) (Fig. 5), we did observe some qualitative differences between the two datasets (Figs. 3*A* and 4*B*) that may relate to individual differences in dynamic brain composition over time.

There is growing evidence that functional connectivity fluctuates over relatively short timescales (i.e., on the order of 0.1–0.01 Hz) (23–25), and as such, it is perhaps unsurprising that similar dynamic patterns exist over longer periods of time. Indeed, it has been hypothesized that such fluctuations in topology are an essential emergent feature of the complex network organization of the brain (26). Evidence for these fluctuations has been shown using computational modeling approaches (27) as well as electrophysiology (28) and more recently, fMRI (11). In addition, other recent work has also shown that alterations in brain network topology track with task performance (29, 30) along with individual differences in intelligence (31) and attentional capacity (32). Together, these findings provide evidence

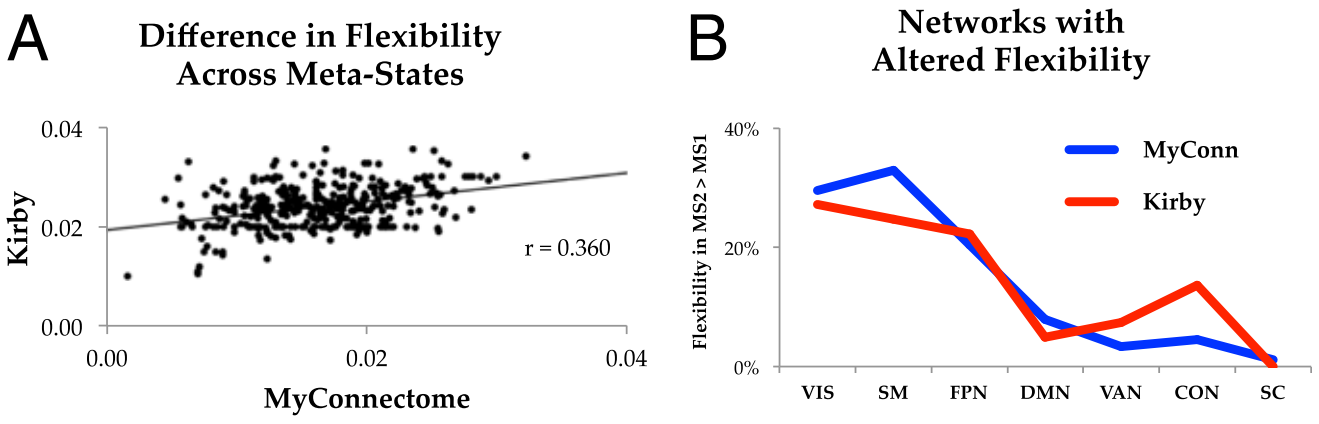

**Fig. 5.** (*A*) Difference in flexibility across the two metastates in both the Kirby (*y* axis) and MyConnectome Project (*x* axis) datasets ($r = 0.360$). (*B*) Similarity of network-level flexibility in both the Kirby (red) and MyConnectome Project (MyConn; blue) datasets—values represent the percentages of significantly different flexibility values that occurred within each of 15 predefined networks (only those with >0 significant regions shown). CON, cingulo-opercular network; DMN, default mode network; FPN, frontoparietal network; SC, subcortical; SM, somatomotor; VAN, ventral attention network; VIS, visual.

for detailed temporal organizational structure within the functional connectome.

An important question facing the field is whether the temporal fluctuations observed in this study vary as a function of neurological and psychiatric disease. Given that impairments in attention are common in neuropsychiatric disorders (33, 34) and that the symptomatology of these conditions often fluctuates over time, it is reasonable to predict that the dynamic interrelationships between large-scale brain systems over longitudinal time might also become impaired in turn. To this end, others have used a similar approach to the one devised in this study to show that dynamic patterns of brain connectivity track with changes in positive mood (35). If this result can be shown to be the case in psychiatric and neurological disorders, the methods described here will have important implications for tracking disease states over time through either the prediction of symptom onset or the response of individual subjects to treatment. In addition, the results will also have an influence on the interpretation of comparisons between clinical groups, wherein differences in network configurations between cohorts may, in fact, reflect differences in temporal dynamics rather than purely spatial pathology per se. Although the path toward solving these issues is currently opaque, it is nonetheless important for studies interrogating brain network abnormalities in cohort studies to broaden their hypothetical lens to include alternative interpretations of significant differences between diseased cohorts.

In conclusion, we have identified a network of cortical and subcortical regions that participate in flexible behavior and alter their time-resolved connectivity profile over longitudinal time, leading to changes in global information processing capacity that track with alterations in self-reported attention. Together, these results support the hypothesis that fluctuations in dynamic interconnectivity between neural regions define the functional capacities of the human brain (6, 36) and also, have important implications for the study of neuropsychiatric disorders that display fluctuations in phenotypic expression of psychological and neurological characteristics over time (8).

## Materials and Methods

Institutional Review Board approval for this study was deemed unnecessary by the University of Texas Office of Research Support.

**ACKNOWLEDGMENTS.** We thank Peter Bell, Krzysztof Gorgolewski, and Craig Moodie for critical insights. The data reported in this paper were made publicly available by the MyConnectome Project, the HCP, and the longitudinal Kirby dataset.

1. Calhoun VD, Miller R, Pearlson G, Adalı T (2014) The chronnectome: Time-varying connectivity networks as the next frontier in fMRI data discovery. *Neuron* 84(2): 262–274.
2. Mantini D, Corbetta M, Romani GL, Orban GA, Vanduffel W (2013) Evolutionarily novel functional networks in the human brain? *J Neurosci* 33(8):3259–3275.
3. Cole MW, Bassett DS, Power JD, Braver TS, Petersen SE (2014) Intrinsic and task-evoked network architectures of the human brain. *Neuron* 83(1):238–251.
4. Zalesky A, Fornito A, Cocchi L, Gollo LL, Breakspear M (2014) Time-resolved resting-state brain networks. *Proc Natl Acad Sci USA* 111(28):10341–10346.
5. Buckner RL, et al. (2009) Cortical hubs revealed by intrinsic functional connectivity: Mapping, assessment of stability, and relation to Alzheimer's disease. *J Neurosci* 29(6): 1860–1873.
6. Bassett DS, Yang M, Wymbs NF, Grafton ST (2015) Learning-induced autonomy of sensorimotor systems. *Nat Neurosci* 18(5):744–751.
7. Barttfeld P, et al. (2015) Signature of consciousness in the dynamics of resting-state brain activity. *Proc Natl Acad Sci USA* 112(3):887–892.
8. Poldrack RA, et al. (2015) Long-term neural and physiological phenotyping of a single human. *Nat Commun* 6:8885.
9. Choe AS, et al. (2015) Reproducibility and temporal structure in weekly resting-state fMRI over a period of 3.5 years. *PLoS One* 10(10):e0140134.
10. Shine JM, et al. (2015) Estimation of dynamic functional connectivity using Multiplication of Temporal Derivatives. *Neuroimage* 122:399–407.
11. Shine JM, et al. (2016) The dynamics of functional brain networks: Integrated network states during cognitive function. arXiv:1511.02976.
12. Bassett DS, et al. (2011) Dynamic reconfiguration of human brain networks during learning. *Proc Natl Acad Sci USA* 108(18):7641–7646.
13. Laumann TO, et al. (2015) Functional system and areal organization of a highly sampled individual human brain. *Neuron* 87(3):657–670.
14. Frey BJ, Dueck D (2007) Clustering by passing messages between data points. *Science* 315(5814):972–976.
15. Rubinov M, Sporns O (2010) Complex network measures of brain connectivity: Uses and interpretations. *Neuroimage* 52(3):1059–1069.
16. Smith SM, et al.; WU-Minn HCP Consortium (2013) Resting-state fMRI in the Human Connectome Project. *Neuroimage* 80:144–168.
17. Hindriks R, et al. (2016) Can sliding-window correlations reveal dynamic functional connectivity in resting-state fMRI? *Neuroimage* 127:242–256.
18. Watson D, Clark LA, Tellegen A (1988) Development and validation of brief measures of positive and negative affect: The PANAS scales. *J Pers Soc Psychol* 54(6):1063–1070.
19. Tagliazucchi E, et al. (2012) Automatic sleep staging using fMRI functional connectivity data. *Neuroimage* 63(1):63–72.
20. Tagliazucchi E, Laufs H (2014) Decoding wakefulness levels from typical fMRI resting-state data reveals reliable drifts between wakefulness and sleep. *Neuron* 82(3): 695–708.
21. Power JD, et al. (2014) Methods to detect, characterize, and remove motion artifact in resting state fMRI. *Neuroimage* 84:320–341.
22. Hutchison RM, et al. (2013) Dynamic functional connectivity: Promise, issues, and interpretations. *Neuroimage* 80:360–378.
23. Raichle ME (2015) The restless brain: How intrinsic activity organizes brain function. *Philos Trans R Soc Lond B Biol Sci* 370(1668):20140172.
24. Allen EA, et al. (2014) Tracking whole-brain connectivity dynamics in the resting state. *Cereb Cortex* 24(3):663–676.
25. Hansen ECA, Battaglia D, Spiegler A, Deco G, Jirsa VK (2015) Functional connectivity dynamics: Modeling the switching behavior of the resting state. *Neuroimage* 105: 525–535.
26. Kelso JA (2012) Multistability and metastability: Understanding dynamic coordination in the brain. *Philos Trans R Soc Lond B Biol Sci* 367(1591):906–918.
27. Deco G, Tononi G, Boly M, Kringelbach ML (2015) Rethinking segregation and integration: Contributions of whole-brain modelling. *Nat Rev Neurosci* 16(7):430–439.
28. Vanrullen R, Busch NA, Drewes J, Dubois J (2011) Ongoing EEG phase as a trial-by-trial predictor of perceptual and attentional variability. *Front Psychol* 2:60.
29. Ekman M, Derrfuss J, Tittgemeyer M (2012) Predicting errors from reconfiguration patterns in human brain networks. *Proc Natl Acad Sci USA* 109(41):16714–16719.
30. Sadaghiani S, Poline J-B, Kleinschmidt A, D'Esposito M (2015) Ongoing dynamics in large-scale functional connectivity predict perception. *Proc Natl Acad Sci USA* 112(27): 8463–8468.
31. Finn ES, et al. (2015) Functional connectome fingerprinting: Identifying individuals using patterns of brain connectivity. *Nat Neurosci* 18(11):1664–1671.
32. Rosenberg MD, et al. (2016) A neuromarker of sustained attention from whole-brain functional connectivity. *Nat Neurosci* 19(1):165–171.
33. Diederich NJ, Goetz CG, Stebbins GT (2005) Repeated visual hallucinations in Parkinson's disease as disturbed external/internal perceptions: Focused review and a new integrative model. *Mov Disord* 20(2):130–140.
34. Allen P, et al. (2012) Neuroimaging auditory hallucinations in schizophrenia: From neuroanatomy to neurochemistry and beyond. *Schizophr Bull* 38(4):695–703.
35. Betzel RF, et al. (2016) A positive mood, a flexible brain. arXiv:1601.07881v1.
36. Davison EN, et al. (2015) Brain network adaptability across task states. *PLOS Comput Biol* 11(1):e1004029.
37. Poldrack RA, et al. (2015) Long-term neural and physiological phenotyping of a single human. *Nat Commun* 6:8885.
38. Shine JM, et al. (2015) Estimation of dynamic functional connectivity using Multiplication of Temporal Derivatives. *NeuroImage* 122:399–407.
39. Mucha PJ, Richardson T, Macon K, Porter MA, Onnela JP (2010) Community structure in time-dependent, multiscale, and multiplex networks. *Science* 328(5980):876–878.
40. Bassett DS, et al. (2011) Dynamic reconfiguration of human brain networks during learning. *Proc Natl Acad Sci USA* 108(18):7641–7646.
41. Guimerà R, Nunes Amaral LA (2005) Functional cartography of complex metabolic networks. *Nature* 433(7028):895–900.
42. Power JD, Schlaggar BL, Lessov-Schlaggar CN, Petersen SE (2013) Evidence for hubs in human functional brain networks. *Neuron* 79(4):798–813.
43. van den Heuvel MP, Sporns O (2013) Network hubs in the human brain. *Trends Cogn Sci* 17(12):683–696.
44. Nichols TE, Holmes AP (2002) Nonparametric permutation tests for functional neuroimaging: A primer with examples. *Hum Brain Mapp* 15(1):1–25.

# Supporting Information

## SI Materials and Methods

**Data and Code Sharing.** All data described here are shared openly at myconnectome.org/wp/data-sharing. Code for all analyses is freely available at https://github.com/macshine/metaconnectivity.

**Data Acquisition.** Functional connectivity was calculated using preprocessed high-resolution resting-state data acquired in 84 unique sessions over the course of 18 mo in a single individual (12). For each session, 10 min of resting-state data were acquired using multiband sequence gradient echo planar image (multiband factor = 8; echo spacing = 0.58 ms; band width = 2,290 Hertz per pixels). The following parameters were used during data acquisition: relaxation time, 1,160 s [echo time (TE) = 30 ms], distance factor = 20%, flip angle was 63° (the Ernst angle for gray matter), field of view was 208 × 180 mm (matrix = 96 × 96), and 2.4 × 2.4 × 2.0-mm voxels with 64 slices.

**Data Preprocessing.** All fMRI data were preprocessed according to a pipeline developed at Washington University (37). First, the initial 100 time points were discarded from the data because of the presence of an evoked auditory signal within the MRI scanner. Data were realigned to correct for head motion and normalized to a mode of 1,000. Each session was registered to a single session that had previously been registered to the mean T1-weighted structural image and atlas. The session to atlas transform was inverted and applied to the mean field map so that the distortion correction could be applied in each session's space. The undistorted data were then reregistered to the atlas space. The transforms for head motion correction and affine registration to atlas space were combined with the field map-based distortion correction to resample the data from the original session space to the undistorted 3-mm isotropic atlas space in a single step using FSLs applywarp tool.

Artifacts were reduced using frame censoring, regression, and spectral filtering. Frames with framewise displacement >0.25 mm were censored as well as uncensored segments of data lasting less than five contiguous volumes (97.1 ± 4% of frames were kept). Nuisance regressors included whole brain, white matter, and ventricular signals and their derivatives in addition to 24 movement regressors derived by Volterra expansion. Interpolation over censored frames was computed by least squares spectral estimation so that continuous data could be band pass-filtered (0.01–0.1 Hz); 12 of 104 sessions were discarded after an alteration to the scanning protocol, and 8 were discarded because of poor normalization.

**Parcellation Scheme.** After preprocessing, the mean time series was extracted from 630 predefined regions of interest (ROIs): 616 cortical parcels (309 from the left hemisphere and 307 from the right hemisphere) from a predefined cortical parcellation scheme (12, 13) that was based on the principles of the Gordon atlas (37) but used all 84 sessions of the individuals' data to create a more subject-specific parcellation scheme along with 14 subcortical parcels from the Harvard–Oxford subcortical atlas (fsl.fmrib.ox.ac.uk/fsl/fslwiki/).

**Session-Specific Resting-State Functional Connectivity.** Using the time series extracted in the previous step, we created a parcel by parcel correlation matrix for each session by calculating a Pearson's correlation between each parcel's time series and then, performing a Fisher's $r$ to $Z$ transformation. We then compared these 84 "session" matrices with one another using spatial Pearson's correlations, leading to the creation of an 84 × 84 matrix that represented the spatial similarity of the correlation matrix for each pair of scanning sessions (Fig. 1*A*).

**Estimation of Temporal Metastates.** We used the affinity propagation technique (14) on the cross-session similarity matrix (Fig. 1*A*) to cluster similar patterns of functional connectivity over time. Briefly, affinity propagation iterates through a similarity matrix and exchanges real-valued "messages" between data points until a high-quality set of exemplars and corresponding clusters emerges. Each scanning session could then be assigned to one of a set of metastates according to the identity of the session with which it was best exemplified. Importantly, this technique does not require the predefinition of the number of clusters, such as is the case with $k$-means clustering and other clustering methods (14); however, there is a relationship between the number of clusters found and the "preference" for when the algorithm should cease "passing messages" between regions. Using this technique with no initial preference toward specific exemplar sessions and the minimum similarity as a constraint on the clustering results (which biases toward the discovery of a small number of clusters), we discovered two clusters of spatially similar matrices in time (Fig. 1*B*)—hereafter referred to as metastate 1 [$MS_1$: 47 sessions; exemplar session 44 (shown in red in Fig. 1*B*)] and metastate 2 [$MS_2$: 37 sessions; exemplar session 83) (shown in blue in Fig. 1*B*)]. We later compared this partition with the results of a clustering based on dynamics (*SI Materials and Methods*, *Estimating the Dynamic Flexibility of Each Brain Region*) and saw broadly similar results.

To ensure that the differences in interareal connectivity were not driven by systematic differences in head motion, we compared the MFD (21) from each session in each metastate using an independent sample $t$ test. In addition, we also compared regional flexibility with the SD of each region's time series and found no significant relationship (mean $r$ = 0.132; $P$ = 0.231), suggesting that regional flexibility was not simply a function of temporal noise. Finally, we also compared modular "switches" (defined as TR to TR mutual information of <0.2) with MFD across 84 sessions and found no significant relationship (mean $r$ = 0.015; $P$ > 0.500). Together, these results suggest that head motion did not adversely affect the interpretations in our manuscript.

**Estimation of Time-Resolved Connectivity.** To estimate functional connectivity between 630 ROIs, we used the MTD technique (Eq. **S1**) (https://github.com/macshine/coupling/) (10). The MTD metric is estimated by calculating the pointwise product of temporal derivative of each pair of regions' time series (38). The MTD tracks similar changes over time, such that a positive value represents "coupling" of time series in the same direction (that is, either both increasing or both decreasing together), whereas negative scores reflect "anticoupling" (that is, one increasing while the other is decreasing). To avoid the contamination of high-frequency noise in the time-resolved connectivity data, the MTD is averaged over a temporal window, $w$:

$$\mathrm{MTD}_{ijt} = \frac{1}{w}\sum_{t}^{t+w}\frac{(dt_{it}\times dt_{jt})}{(\sigma_{dt_i}\times\sigma_{dt_j})} \quad \textbf{[S1]}$$

Eq. **S1** shows MTD. For each time point, $t$, the MTD for the pairwise interaction between regions $i$ and $j$ is defined according to Eq. **S1**, where $dt$ is the first temporal derivative of the $i$th or $j$th time series, σ is the SD of the temporal derivative time series for region $i$ or $j$, and $w$ is the window length of the simple moving average. This equation can then be calculated over the course of a time series to obtain an estimate of time-resolved connectivity between pairs of regions.

Previous work based on simulated blood oxygen level dependent (BOLD) data has shown that a window length of seven TRs provides optimal sensitivity and specificity for detecting dynamic changes in functional connectivity structure when using the MTD technique (10). Given that we used a 0.1-Hz low-pass filter on our data, in theory, all signals with periods of 10 s or smaller should be removed from the data. As such, we opted to use a temporal window of 10 time points to calculate a simple moving average of the MTD. The MTD can then be tracked over time to provide an estimate of time-resolved functional connectivity, because within each window, the value of the MTD is directly interpretable as the extent to which two regions showed similar changes in activity over time. Using the MTD value, we were, thus, able to compute a weighted and signed adjacency matrix within each temporal window.

**Time-Resolved Community Structure.** To determine the presence of community structure that existed over the temporal domain, we used a version of the Louvain algorithm (Eq. **S2**) to track community structure over time during each resting-state session (39). Briefly, the Louvain algorithm iteratively maximizes the modularity statistic, $Q$, for different community assignments until the maximum possible score of $Q$ has been obtained. The modularity estimate for a given network is, therefore, a quantification of the extent to which the network may be subdivided into communities with stronger within- than between-module connections. This algorithm optimized the modularity quality function, $Q$, by identifying community structure in networks that are "linked" together in time. In line with previous work (6), the γ- and ω-parameters, which index the expected size of communities and the strength of links between modules in time, respectively, were each set to one. In keeping with previous studies, we reanalyzed our data across a range of γ- and ω-parameter spaces (0.9–1.1 in steps of 0.05) and found robustly similar module structure (mean mutual information across sessions: 0.739 ± 0.18 across the parameter space). Finally, because of the stochastic nature of the modularity maximization algorithm, we estimated the community structure of the data 100 times for each session and then, estimated a consensus clustering for each temporal window:

$$Q=\frac{1}{2\mu}\sum_{ijlr}\left\{\left(A_{ijl}-\gamma_l N_{ijl}\right)\gamma_{lr}+\delta_{ij}\omega_{jlr}\right\}\delta\left(M_{il},M_{jr}\right) \qquad \textbf{[S2]}$$

Eq. **S2** shows the multiscale, multiscale modularity algorithm (10), where $A_{ij}$ is the adjacency matrix, $N_{ijl}$ represents the Newman–Girvan null model (2), $\gamma_l$ is the structural resolution parameter of layer $l$, the parameter-$\omega_{jlr}$ is the "interlayer coupling parameter" between node $j$ in layer $r$ and node $j$ in layer $l$, and $\delta_{MilMjr}$ is set to one when regions are in the same community and zero otherwise. In keeping with previous works (2, 16), we set $\omega = 1$ and $\gamma = 1$.

**Estimating the Dynamic Flexibility of Each Brain Region.** In keeping with the work by Bassett et al. (40), we estimated the flexibility of each brain parcel by calculating the percentage of temporal windows in which an individual region "switched" between modules normalized to the total number of modules in the data (as estimated in the previous step). Accordingly, regions with high flexibility sampled from a wider range of the total brain than those with low flexibility, which were more regularly present in a temporal community with a smaller proportion of the total brain. Code was obtained directly from the original author (www.danisbassett.com/resources.html). To assess this measure across a large dataset, we first calculated the flexibility of each of 333 cortical and 14 subcortical parcels in an independent cohort of 100 unrelated subjects from the HCP dataset (16). Because of a shorter repetition time (TR = 0.72 s), a sliding window of 14 was used to calculate the simple moving average of the MTD. The two estimates of flexibility during rest showed moderate spatial correspondence ($r = 0.440$).

To test for significance between the identified metastates, the flexibility of each region was compared between the two metastates using an independent samples $t$ test with a false discovery rate of $\alpha = 0.05$ applied to provide control over multiple comparisons. The spatial projection was then mapped onto a 32,492-vertex surface rendering using the Connectome Workbench (www.humanconnectome.org/).

**Cartographic Profiling.** Based on time-resolved community assignments, we estimated within-module connectivity by calculating the time-resolved module-degree $Z$ score ($W_T$; within-module strength) for each region in our analysis (Eq. **S3**) (41):

$$W_{iT}=\frac{\kappa_{iT}-\bar{\kappa}_{s_{iT}}}{\sigma_{\kappa_{s_{iT}}}} \qquad \textbf{[S3]}$$

Eq. **S3** shows the module-degree $Z$ score, $W_{iT}$, where $\kappa_{iT}$ is the strength of the connections of region $i$ to other regions in its module $s_i$ at time $T$, $\bar{\kappa}_{s_{iT}}$ is the average of κ over all of the regions in $s_i$ at time $T$, and $\sigma_{\kappa_{s_{iT}}}$ is the SD of κ in $s_i$ at time $T$.

The participation coefficient, $B_T$, quantifies the extent to which a region connects across all modules (i.e., between-module strength) and has previously been used to characterize hubs within brain networks (42, 43). The $B_T$ for each region was calculated within each temporal window using Eq. **S4**:

$$B_{iT}=1-\sum_{s=1}\left(\frac{\kappa_{isT}}{\kappa_{iT}}\right)^2 \qquad \textbf{[S4]}$$

Eq. **S4** shows participation coefficient $B_{iT}$, where $\kappa_{isT}$ is the strength of the positive connections of region $i$ to regions in module $s$ at time $T$, and $\kappa_{iT}$ is the sum of strengths of all positive connections of region $i$ at time $T$. The participation coefficient of a region is, therefore, close to one if its connections are uniformly distributed among all of the modules and zero if all of its links are within its own module.

Using these two values for each region within each temporal window, we were able to estimate the cartographic profile for the entire brain over time. Specifically, we created a joint histogram of each temporal window (which is naïve to cartographic boundaries) by summing the instances of each value of $W_T$ and $B_T$ within 100 equally defined bins along each axis. To compare the cartographic profile of the two metastates, we compared the intensity within each bin of the joint histogram between the two metastates using an independent samples $t$ test (FDR $P < 0.05$).

**Efficiency of Communication Within Metastates.** To determine the mean global efficiency of each metastate, the time-averaged connectivity matrix for each session was thresholded to retain the top 5% of positive connections (although results were consistent across a range of thresholds: 1 to 20%). We then calculated the global efficiency of the network ($E$) using Eq. **S5**. Importantly, the efficiency was calculated on the largest connected component in the data, ignoring nodes that were "isolated" from the rest of the graph. The largest component in the data after thresholding the top 5% of the data contained 93.54 ± 0.02% of 630 regions on average, suggesting that the presence of isolated nodes was relatively rare (6.46 ± 0.02% on average).

These values were compared between metastates using two independent samples $t$ tests (an FDR of $\alpha = 0.01$ was used for the local efficiency results):

$$E = \frac{2}{n(n=1)} \sum_{i<j \in G}^{n} \frac{1}{d(i,j)} \qquad \textbf{[S5]}$$

Eq. **S5** shows global efficiency of a network, where $n$ denotes the total nodes in the largest connected component, and $d(i,j)$ denotes the shortest path between node $i$ and neighboring node $j$.

**Relationship to Self-Reported Behavior.** To determine whether the states identified in the neuroimaging sessions were related to behavior, results from the self-reported Positive and Negative Affect Schedule (18) questionnaires were compared between the two sessions using a series of Mann–Whitney $u$ tests (FDR $\alpha = 0.01$). Scores from the significant questions fell into two related groups: one associated with a higher frequency of response in metastate 1 (fatigue questions: drowsy: $Q_{28}$; sleepy: $Q_{57}$; sluggish: $Q_{58}$; and tired: $Q_{62}$) and another associated with a higher frequency of response in metastate 2 (attention questions: attentive: $Q_{11}$; concentrating: $Q_{18}$; and lively: $Q_{43}$). These scores were then compared on days associated with fasting and caffeine to determine whether the fluctuations in self-reported behavior tracked with this particular behavioral manipulation or rather, a broader organization.

**Vector Autoregressive Null Model.** To determine whether the clustering identified in the longitudinal data could be observed in truly stationary data, we simulated data using a series of stationary, 2D vector autoregressive (VAR) models, which were used to generate surrogate regional time series satisfying the null hypothesis of a linearly correlated, stationary, multivariate stochastic process. Consistent with the approach adopted by Zalesky et al. (4), the VAR model order was chosen to minimize the Bayesian information criterion (BIC), which was evaluated over a range of model orders 1–20 for 630 pairs of regions used in this study. The BIC was minimized for a model order of six, which is roughly the predicted time associated with the peak of BOLD response (and as such, is consistent with the results obtained in ref. 4).

The mean covariance matrix across all 84 sessions from the discovery group was used to generate 2,500 independent null datasets, which allows for the appropriate estimation of the tails of nonparametric distributions (44). These time series were then filtered in a similar fashion to the BOLD data and used to estimate the time-averaged connectivity of each session. For each of 2,500 iterations, the affinity propagation clustering technique was used to cluster the 84 × 84 matrix comparing the time-averaged connectivity pattern of each simulated session. In each of 2,500 iterations, the clustering method only identified a single cluster, suggesting that the presence of two clusters was significantly greater than would be expected by chance if there was no longitudinal fluctuation in functional connectivity.

**Phase Randomized Null Model.** To ensure that the major outcome measures in our study were not caused by overestimation of nonsignificant fluctuations in the data, we ran a second null model, in which we coherently randomized the Fourier phase of the data. This technique was applied to the time series from each session; after applying our analysis pipeline to these null data, each of the major outcome measures in the study (i.e., flexibility, cartography, and efficiency) was estimated, and separate null distributions were created by taking the maximum difference between metastate 1 and metastate 2 for each outcome measure across 2,500 iterations. Values more extreme than the 95th percentile of this distribution were, thus, considered to be significant (44).

**Reproducibility of Results in an Independent Longitudinal Dataset.** To reproduce our results, we collected open source rfMRI data from a separate dataset (https://www.nitrc.org/projects/kirbyweekly/). Although conceptually similar to the project described above, these data were collected on a different scanner (3T Philips Achieva Scanner), and there were some differences in the rfMRI data, which were acquired using a multislice sensitivity encoded-echo planar image (SENSE-EPI) pulse sequence with TR/TE = 2,000/30 ms, SENSE factor = 2, flip angle = 75°, 37 axial slices, nominal resolution = 3 × 3 × 3 $mm^3$, 1-mm gap, 16-channel neurovascular coil, and number of frames per run = 200. Data were extracted from 333 cortical regions [from the atlas by Gordon et al. (37), which used a similar technique to the individual analysis] and 14 subcortical regions (fsl.fmrib.ox.ac.uk/fsl/fslwiki/). Because of differences in spatial coverage between the group- and individual-level parcellations, we recalculated the flexibility of each brain region from the individual subject sessions using the 333-region parcellation scheme. These data were preprocessed and analyzed in an identical fashion to the aforementioned study; however, because of the longer repetition time used in the replication sample (TR = 2 s), a window length of five time points (∼10 s) was used to calculate the simple moving average of the MTD.